\documentclass{ws-procs9x6}
\usepackage{picinpar}
\begin{document}

\title{Quarkonium at finite temperature
\footnote{\uppercase{B}ased on work done in 
collaboration with \uppercase{S}. \uppercase{D}atta, \uppercase{F}. 
\uppercase{K}arsch and
\uppercase{I}. \uppercase{W}etzorke.  }}
\author{P. Petreczky
\footnote{\uppercase{G}oldhaber fellow, supported  
under contract 
\uppercase{DE}-\uppercase{AC}02-98\uppercase{CH}10886 with 
the \uppercase{U.S.} \uppercase{D}epartment of \uppercase{E}nergy.}}
\address{Physics Department, Brookhaven National Laboratory,
Upton, NY 11973, USA}


\maketitle
\vspace*{-0.4cm}
\abstracts{
I discuss quarkonium spectral functions at finite temperature
reconstructed using the Maximum Entropy Method.
I show in particular that the $J/\psi$ survives
in the deconfined phase up to $1.5T_c$}
\vspace*{-0.3cm}
The study of quarkonium system at finite temperature 
has been a subject of considerable interest since
the work of Matusi and Satz \cite{ms}, but a first principle
calculation of quarkonium properties at non-zero 
temperature was missing. It was
shown, however, that the application of the Maximum Entropy
Method (MEM) can make such calculation possible \cite{mem}. The method
have been successfully applied at zero \cite{mem} as well as at
finite temperature \cite{ft}.
\begin{window}[0,r,
{\includegraphics[width=2.1in]{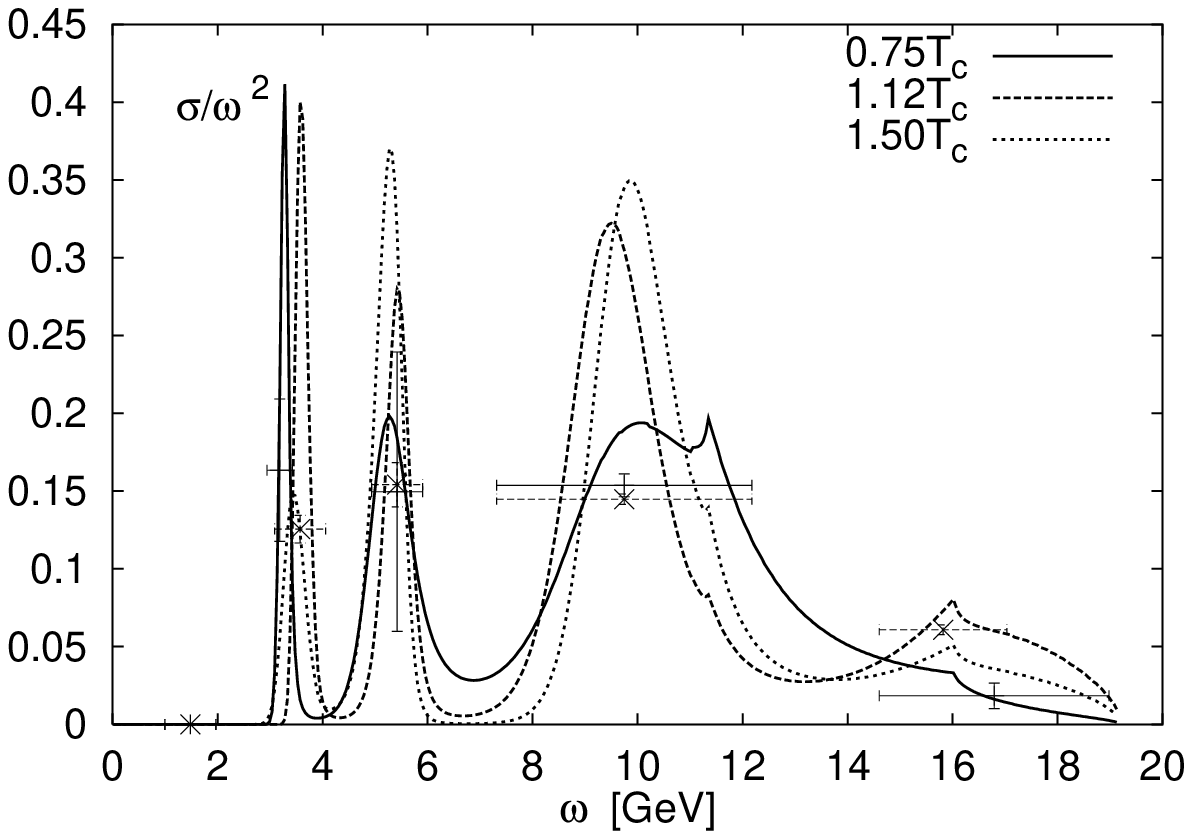}},{}]
I am going to discuss charmonium
spectral function calculated with MEM on $48^3 \times N_{\tau}$
lattices at lattice spacing $a^{-1}=4.86$GeV and $N_{\tau}=24,16$
and $12$ corresponding to temperatures $0.75T_c$, $1.12T_c$
and $1.5T_c$ ($T_c$ being the deconfinement temperature). 
The results for the vector channel are shown
in the Figure. As one can see the $J/\psi$ seems to survive
up to temperatures $1.5T_c$. Similar calculation have 
been performed in the scalar and axial vector channels
which correspond to the $P$-state charmonia, but no peak
was found there.
\end{window}

\end{document}